\documentclass[preprint]{aastex}

\usepackage{aas_macros} \usepackage{graphicx}

\begin{document}

% ************* Title and Metadata-Stuff ***************

\title{Nonlinear Wave Interactions as Emission Process of Type II Radio Bursts}

\newcommand{\etal}{\MakeLowercase{\textit{et al. }}} % "et al."

\author{Urs Ganse}
\affil{Lehrstuhl f\"ur Astronomie, Universit\"at W\"urzburg}
\email{uganse@astro.uni-wuerzburg.de}
\author{Patrick Kilian}
\affil{Lehrstuhl f\"ur Astronomie, Universit\"at W\"urzburg}
\author{Felix Spanier} \affil{Lehrstuhl
f\"ur Astronomie, Universit\"at W\"urzburg}
\author{Rami Vainio}
\affil{Department of Physics, University of Helsinki}

% ***************** Abstract *******************

\begin{abstract}
The emission of fundamental and harmonic frequency radio waves of type II radio
bursts are assumed to be products of three-wave interaction processes of
beam-excited Langmuir waves. Using a particle-in-cell code, we have performed
simulations of the assumed emission region, a CME foreshock with two
counterstreaming electron beams.
Analysis of wavemodes within the simulation shows self-consistent excitation of
beam driven modes, which yield interaction products at both fundamental and
harmonic emission frequencies.
Through variation of the beam strength, we have investigated the dependence of
energy transfer into electrostatic and electromagnetic modes, confirming the
quadratic dependence of electromagnetic emission on electron beam strength.
\end{abstract}


\maketitle

% ***************** Text text text! *****************
\section{Overview}

Emission of radio bursts from the solar corona has been observed since
the early 1950ies \citep{WildMcCready}. While theoretical explanations for the
emission mechanisms of most types of bursts have been found, many details of
the emission process of type II radio bursts are still a matter of discussion.

The morphology of type II bursts typically shows a two-band emission
spectrum, consisting of the \emph{fundamental} emission band and of the
\emph{harmonic} emission band at about twice the frequency of the fundamental
\citep[in a few, near-limb events signals of third harmonic emission are also
discernable, see][]{ThirdHarmonic}. The fundamental frequency is believed
to correspond to the plasma frequency of the emission region, slowly decreasing
over time as the coronal/interplanetary shock travels outwards into the heliosphere.
\citep{Cane1987, NelsonMelrose}

The commonly accepted model by \cite{reiner1998} proposes that
the emission region is located around points where the CME drives a curved
perpendicular shock, leading to efficient shock drift acceleration of
electrons. Due to the curved nature of the shock, these can escape into the
foreshock following the magnetic field, where they form an electron beam
population.
This model has been thoroughly treated analytically. \cite{KnockModel} have
derived theoretical emissivities for Langmuir wave emission and subsequent
conversion into electromagnetic emissions, and \cite{schmidtCMEshocks} have
combined these with MHD simulations to track the radio burst emission region as
it moves outward from the Sun.
These theoretical predictions are however not in complete agreement with
observations, and many details of the fine structure (band splitting and
herringbone patterns) often visible on type II bursts are yet to be
explained\citep{Aurass94}.

Similar to the assumed emission process of type III bursts, instabilities
driven by the electron beams can then lead to creation of Langmuir waves $(L)$
and their three wave interaction products (Ion-soundwaves $S$ and transverse
electromagnetic waves $T$) \citep{Melrose}:
\begin{eqnarray}
	L &\rightarrow& L' + S \label{eq1} \\
	L &\rightarrow& S + T(\omega_{pe}) \label{lts}\\
	L + L' &\rightarrow&T(2\omega_{pe}) \label{llt} \label{melrosecouplings}
\end{eqnarray}

Momentum and energy conservation as well as wave polarization impose certain limits on the waves' $\vec{k}$
directions \citep{SpanierVainio09}. The Langmuir waves (L) should primarily be excited parallel to the
beam direction, potentially following the beam electrons' pitch angle distribution
\citep{KarlickyVandas}, whereas electromagnetic waves are expected to be emitted
in quasi-perpendicular directions: Assuming that
\begin{eqnarray}
  k^L_\parallel &=& k_\parallel^T + k_\parallel^S
	\label{analytic1}\\
	k^L_\perp &=& k_\perp^T + k_\perp^S\\
	\omega^L &=&	\omega^T + \omega^S,
  \label{analytic2}
\end{eqnarray}

Angular momentum conservation leads to polarization selection
rules \citep{SpanierVainio09}. Inserting the dispersion relations for all three
waves and using these selection rules to neglect $k_\parallel^T$ as well as $k_\perp^S$, we get
\begin{equation}
	3k_\parallel^2 v^2_\mathrm{th}
		= c^2 k_\perp^2 +	k_\parallel^2 v_s^2 + 2k_\parallel v_s \omega_{pe} + k_\perp k_\parallel v_s	 c,
		\label{analytic3}
\end{equation}
(with speed of sound $v_s$ and electron thermal speed $v_\mathrm{th}$) which
results in an expected $k$-value for the harmonic electromagnetic emission of
\begin{equation}
	k_\perp = -\frac{v_s}{2c} k_\parallel \pm \sqrt{3k_\parallel^2 \left(
	\frac{v_\mathrm{th}^2}{c^2} - \frac{v_s^2}{4 c^2} \right) - 2
	\frac{v_s}{c^2}k_\parallel \omega_{pe}}.
	\label{analytic4}
\end{equation}
\label{theory}
Being a second-order coupling of Langmuir waves, the intensity of the
electromagnetic emission is expected to be quadratic in the Langmuir wave
energy, and should thus also be quadratic in the initial electron beam energy
\citep{KnockModel} for the model described here. 
This assumes a fixed level of ion soundwave intensity, which the simulations in
this work provide through identical thermal initialization of the background
ion population in all runs.  Excitation of additional soundwave intensity by
the beam can be neglected due to the small simulation timescales.

\medskip
In the simplest case of this emission scenario, a single electron beam emission
site, created on a smoothly curved CME shock, is supplying the electron beam
populations for radio emission.  Langmuir waves excited through the beam driven
instability would first have to
scatter off soundwaves (Eq. \ref{eq1}) at a sufficient rate to create a
suitable population of counter-propagating Langmuir waves $L'$. If this process
is too slow, harmonic emission (Eq. \ref{llt}) cannot be expected to attain
intensities which are comparable to the fundamental plasma emission process.

An alternative scenario can be constructed by assuming not one single, but two
(or multiple) points along a shock front in which perpendicular shock
conditions pertain, allowing for drift acceleration of electrons into the
foreshock. This could be caused by shock ripples at the front of the CME, or
density fluctuations in the background solar wind leading to inhomogeneous
propagation of the shock through the heliosphere.
With multiple electron acceleration sites feeding electron beams into the
foreshock from different directions, counterstreaming electron beams may be
present within the emission region. Direct excitation of counter-propagating
Langmuir waves through beam driven instability is hence possible, leading to a
stronger harmonic emissivity.

\medskip

Due to the relatively small size of the emission region, in-situ satellite
observation data is very scarce. In a handful of cases however, measurements
of interplanetary type II burst emission regions were possible, as an
outward-expanding CME shock front encountered spacecraft at 1AU\citep{Pulupa2007}.
The measurements obtained from these emission-region crossings agree with the
proposed model, in that they provide evidence of counterstreaming foreshock electron beam
presence. Electric field measurements on board the spacecraft are also
indicating a strong presence of waves near the plasma frequency, attributed to
Langmuir waves.
However, direct in-situ satellite measurements of wave quantities are limited
to the resolution of $\omega_s$ in the spacecraft's frame. Since the solar wind is
streaming by the spacecraft at a velocity $\vec{u}_{sw}$, this quantity is a
mixture of $\omega_s = \omega + \vec{u}_{sw} \cdot \vec{k}$, but the $k$
contribution is negligible for the high-frequency wavemodes of interest to this
study. While multispacecraft observations have been successful in
reconstructing directionalities of radio bursts \citep{Thejappa2012}, a detailed
analysis of the kinematics of three wave interaction processes, for which
information with both $\omega$ and $\vec{k}$ resolution is essential, can only
be provided by simulations.

\subsection{Related Work}
In similar work, \cite{KarlickyVandas} performed 1.5D simulations of
beam-excitation of Langmuir waves from foreshock electrons, and subsequent
couplings to kinetic wavemodes. Due to the limitation to one spatial dimension,
these runs were not able to represent waves with transverse momentum and hence
precluded the subsequent electromagnetic emission mechanism (eqs. \ref{lts}, \ref{llt}).
In \cite{KarlickyIAUS}, 2.5D PiC simulations were conducted into which
monochromatic, coherent Langmuir waves were artificially injected as an initial
condition instead of an electron beam, yielding a strong signal of both
fundamental and harmonic emission.
\cite{Guangli} focused on the generation of backward-propagating waves through
scattering (Eq.\ref{eq1}), confirming that direct creation of these waves is
difficult in a particle-in-cell simulation.
Finally, the simulations of \cite{SakaiBullshit} attempt to model the shock
structure itself within the simulation box, in an attempt to represent the
complete emission mechanism within one self-contained simulation. Due to the
inherent numerical limits given by a PiC code, both the accelerated particles
and wave quantities are at odds with the other papers presented here.

\section{Simulation}

\begin{figure*}[bht] \centering
	\includegraphics[width=15.5cm]{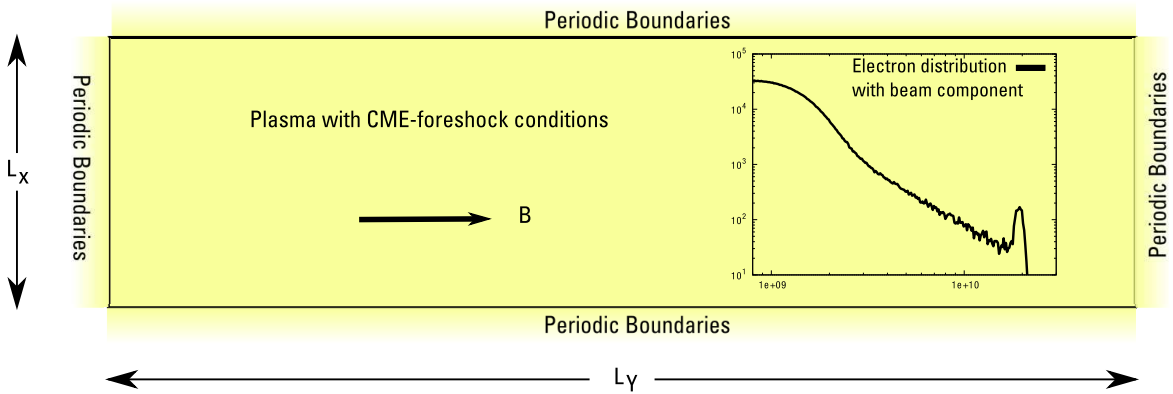} \caption{Numerical
	simulation setup of the 2.5D PiC simulations. In addition to the homogeneous,
	thermal background plasma, two counterstreaming electron beam populations are
	present.} \label{setup} \end{figure*}

Since the phenomenology of beam-driven wave instabilities depends on an electron
distribution function that is far from equilibrium, MHD or fluid simulations
are not suitable for this problem, but rather kinetic simulation methods have
to be employed, with correspondingly larger computational requirements.
To obtain and analyze full spatial information about wave processes within
the CME foreshock region (as opposed to 1D satellite measurements), the kinetic
particle-in-cell simulation code ACRONYM \citep{acronym11} was used.
This code, developed and maintained at the Department of Astronomy, University
of W\"urzburg, is a fully relativistic, 2nd order particle-in-cell code for
astronomical, heliospheric and laboratory plasmas. Using MPI-parallelization,
the code is running on all major supercomputing platforms.

Due to the inherent length- and timescale requirements (governed by the plasma
Debye length of about $\lambda_D \approx 1\,\mathrm{cm}$) of kinetic
simulations,
it is impossible to model a complete CME, or even a significant part of the
actual shock front within the bounds of a simulation. Rather, the microphysics
of plasma wave interaction within the foreshock region is of central importance
here. The actual simulations therefore model only the foreshock region,
with the electron beam initialized as external input.

Improving on earlier work \citep{GanseIAUS}, the focus of this paper is the
self-consistent creation of electromagnetic emission from Langmuir waves which
have not been artificially created as monochromatic external input, but excited
through instabilities within the simulation itself. Correspondingly larger
timescales and better spectral resolution are required over
\cite{KarlickyIAUS}.
The simulations are set up as 2.5D rectangular grids with periodic boundaries,
which are homogeneously filled with the background foreshock plasma under
quiescent coronal conditions ($T\approx 1 \,\mathrm{MK}$, $\rho=2.5\cdot10^{7}
\mathrm{cm}^{-3}$, $B=1 \mathrm{G}$), with a thermal particle distribution. On
top of that, two counterstreaming beamed electron populations are added at
$v_\mathrm{Beam} \approx 5 v_\mathrm{th}$, whose density is $10\%-15\%$ of the
total electron density (see Fig. \ref{setup}), based on in-situ observation data \citep{KnockModel}. Their pitch angle distribution
is centered around $45^{\circ}$, following \cite{KarlickyVandas}.

The selection of two counterstreaming electron populations as opposed to a
single beam is motivated by our earlier results
\citep{UrsSolarPhysics}, as well as work by \cite{Guangli}, showing
that a creation of backward-propagating Langmuir waves of sufficient strength
within a PiC simulation is itself numerically challenging. The counterstreaming
beam setup provides symmetric creation of forward- and backward-propagating
waves, and can physically occur in scenarios of multiple shock drift
acceleration sites \citep{keyhere}.

To obtain physically valid results, the electron to proton mass ratio has been
chosen as the physical $m_p/m_e = 1836$, the protons being part of the
homogeneous, thermal background.
In order to reduce the fundamental problem of numerical noise in our
PiC-simulations, the particle number per cell had to be chosen quite large.
Results have shown that below 100 particles per cell, nonlinear wave
interactions were strongly disturbed by nonphysical effects due to statistical
noise. This also sets a lower limit to the beam intensities which can be
simulated given the current computational resources, as the generated wave
signals have to be sufficiently stronger than the noise background. They are however still in reasonable agreement with \cite{KnockModel}.

Using large supercomputers, numerical simulation extents of $8192\times4096$
cells with $200$ particles per cell were attainable, leading to physical
extents of $80.28 \times 40.14$ metres. While this is still many orders of
magnitude smaller than the expected size of the complete emission region, it
resolves the kinetic plasma length scales and the wavelengths of all
contributing waves with sufficient headroom.

\begin{figure}[htb]
	\begin{center}
		\rotatebox{270}{\includegraphics[height=\textwidth]{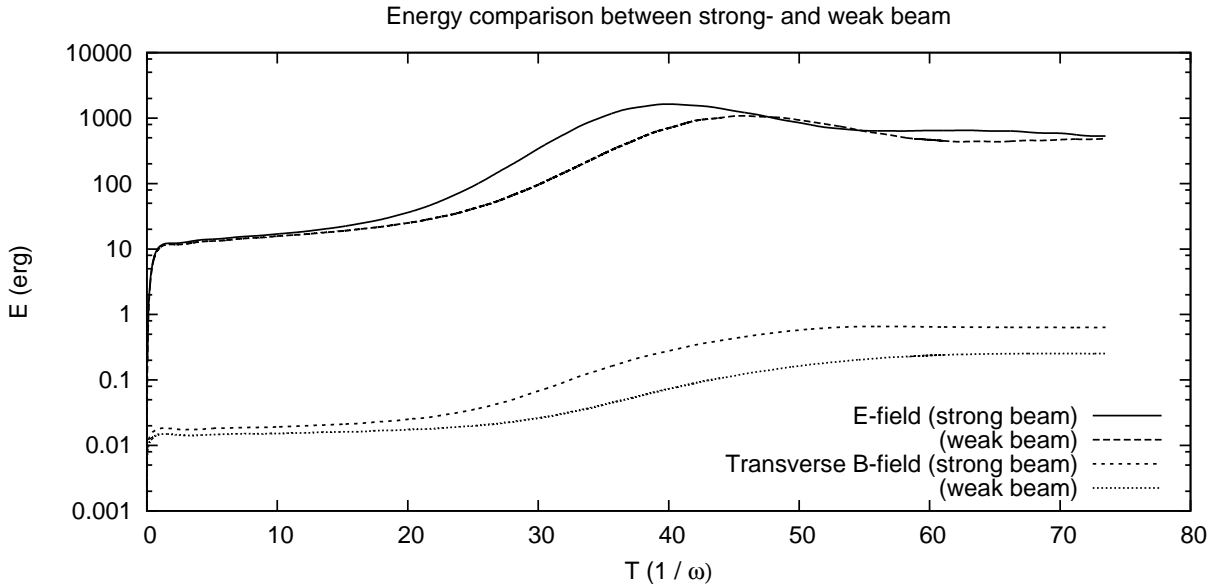}}
	\end{center}
	\caption{Energy distribution in two foreshock electron beam simulations. A
	transfer of energy into electric fields and subsequently into transverse
	magnetic fields is taking place, starting from $t\sim20\omega_{pe}^{-1}$.
	In a direct comparison of two simulations with a $50\%$ difference in
	electron beam density, a $52.7\%$ increase in peak electric field intensity
	is observed. The resulting gain in magnetic field energy exceeds that,
	yielding a $172\%$ increase.}
	\label{fig:energyout}
\end{figure}

Multiple simulation runs were conducted to study the influence of multiple
input parameters on the wave interaction behaviour in this scenario: The basic
simulation setup outlined above was varied in beam density (a $50\%$ increase)
and background plasma density. 

\section{Results}

In the following, the results of two simulation runs, one with electron beam
density $10\%$ of the background plasma density, and one with $15\%$ of
background plasma density shall be presented and compared.
Firstly, the spatially averaged distribution of energy into particle kinetic
energy, electric and magnetic fields will be discussed, followed by
decomposition of the field's energy contributions into individual spatial modes.

The energy distribution within the simulations' constituents (Fig. \ref{fig:energyout})
yields that the initial energy, stored only in the particles' motion and the
background B-field, starts to convert into electric field energy at around
$t = 20 \omega_{pe}^{-1}$. Soon afterwards, energy conversion into transverse
B-field components is also observed.

The initial transfer of energy into electric fields is attributed to the
excitation of Langmuir waves by the presence of the electron beam, consistent
with theoretical predictions. The peak energy content in the electric field is
about $2\%$ of the initial electron beam's kinetic energy, confirming the results of
\cite{KarlickyVandas}.
The conversion of electric into magnetic energy (in the transverse components
of the B-field) is then attributed to the creation of transverse
electromagnetic waves.

A direct comparison between the simulation with the weaker electron beams
(electron density $n_\mathrm{beam} \sim 10\%\; n_\mathrm{bg}$) and the
stronger-beam simulation ($n_\mathrm{beam} \sim 15\%\; n_\mathrm{bg}$, same beam
velocity), given in Table \ref{tab:zahlen} shows that the increase in beam energy leads to a proportionate
increase in peak electric field strength, and a much larger increase in
transverse magnetic field energy.
The increase in electric field energy by $53\%$ is consistent with the linear
relationship of beam strength to electrostatic mode excitation by
Cherenkov-type instabilities.

\begin{table}[thb]
\begin{center}
\begin{tabular}{ccc}
	
	\hline
	\hline
	\parbox{5cm}{Beam electron density\\ (as $\%$ of $n_\mathrm{bg}$)} & \parbox{5cm}{Peak electric field energy} & \parbox{5cm}{Peak transverse\\ mag. field energy}\\
	\hline
	10 &  $1082\,\mathrm{erg}$ &  $0.18\,\mathrm{erg}$\\
	15 &  $1652\,\mathrm{erg}$ &  $0.49\,\mathrm{erg}$\\
	\hline
	Energy ratio & 1.53               &      2.72\\
	\hline
	\hline
\end{tabular}
\end{center}
\label{tab:zahlen}
\caption{Comparison of two simulation runs, in terms of electron beam intensity and resulting energy content in electric and transverse magnetic field components.}
\end{table}

If the transverse magnetic field strength is actually caused by second-order
wave couplings of electrostatic wavemodes, as outlined in section \ref{theory},
a quadratic increase in energy in this component relative to the electrostatic
mode energy is expected: The observed increase by a factor of $2.72$ is in
reasonable agreement with this prediction, as $\sqrt{2.72} = 1.65$ is
comparable to the growth of the electrostatic mode by a factor of $1.53$.
Higher-order nonlinear effects might be the cause for additional energy excess
here.

In order to obtain more detailed quantitative information about the energy
distribution into different wave modes within the foreshock plasma, the
discretized simulation data was Fourier-transformed both in space, allowing
analysis of instantaneous k-space spectra, and in time, to obtain dispersion
diagrams.

\begin{figure}[htp]
	\begin{center}
		\rotatebox{270}{\includegraphics[height=8cm]{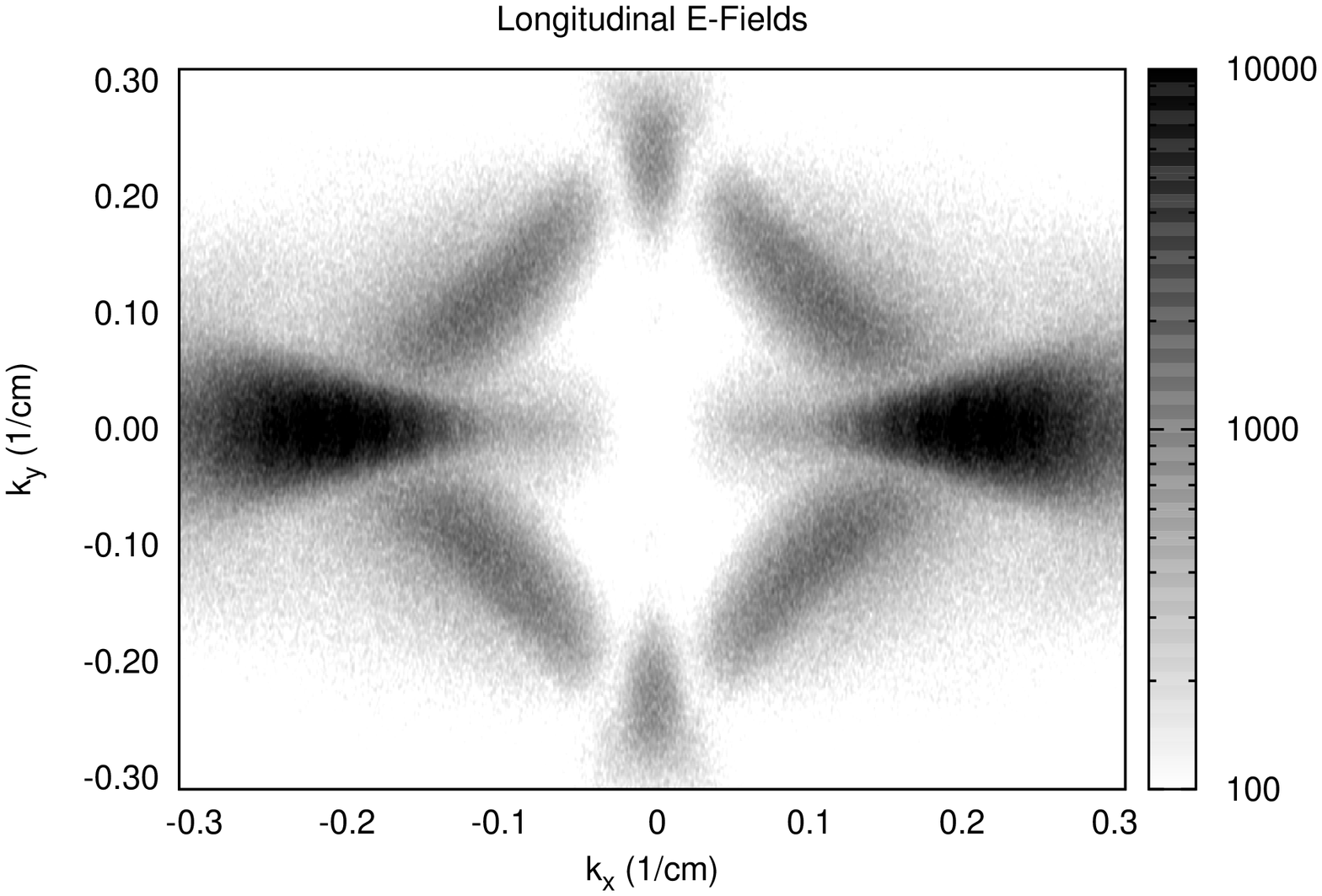}}
		\rotatebox{270}{\includegraphics[height=8cm]{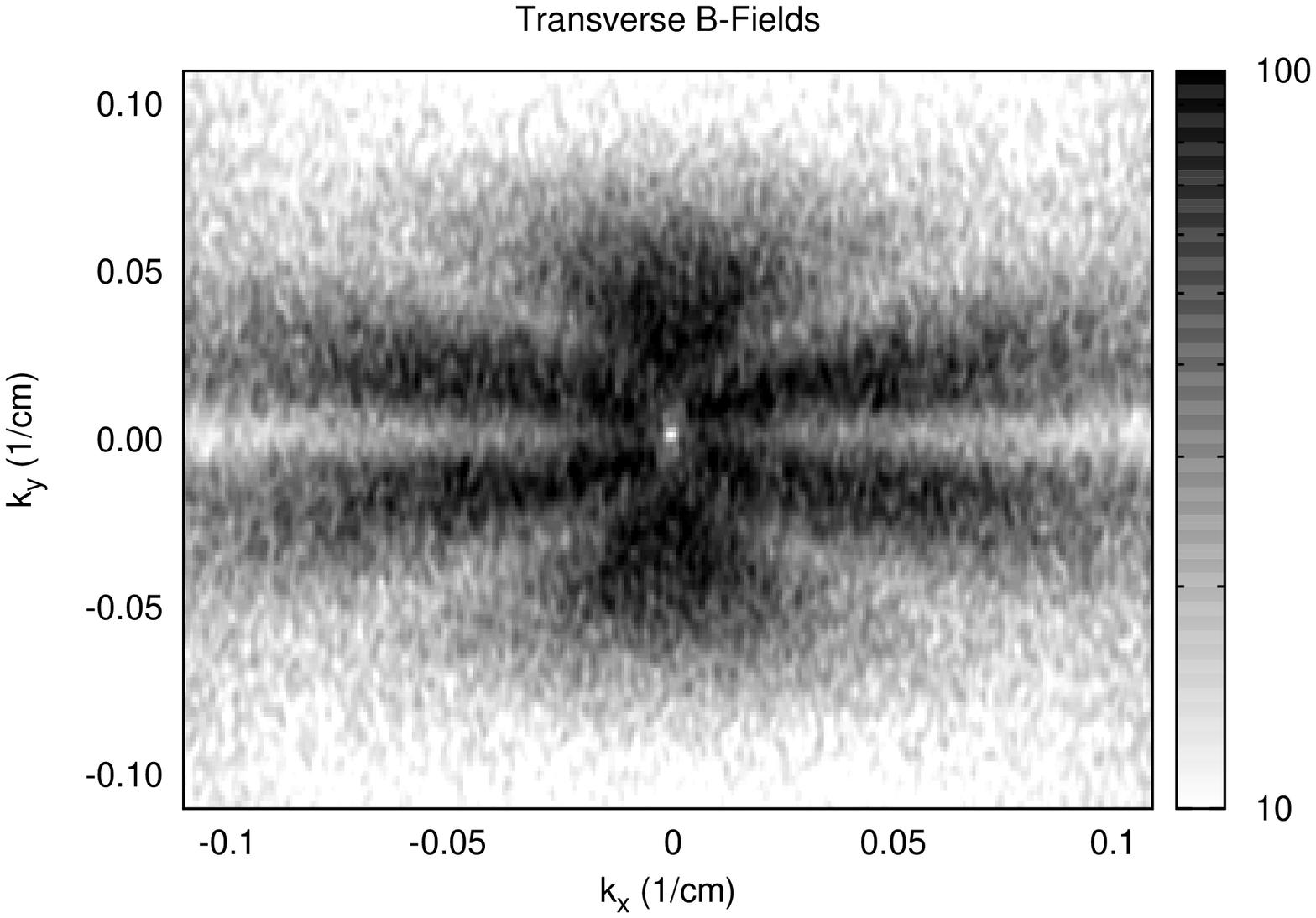}} 
	\end{center}
	\caption{$k$-space distribution of wave energies in the foreshock plasma
	(arbitrary units).  \textbf{Left:} Longitudinal electric field,
	\textbf{Right:} Transverse magnetic field. The x-axis corresponds to the
	electron beam direction. } 
	\label{fig:butterfly} 
\end{figure}

Figure \ref{fig:butterfly} shows $k$-space intensities of longitudinal $E$- and
transverse $B$-Fields at $t= 65\,\omega_{pe}^{-1}$. The $x$-axis corresponds to the
electron streaming direction.

By taking the Fourier transform in both temporal and spatial dimensions, the
dispersion plots in Fig. \ref{fig:dispersione} and \ref{fig:dispersionb} are
obtained. They respectively show the relative intensities of the longitudinal
electric and transverse magnetic fields in relation to $\omega$ and $k$.

\subsection{Langmuir wave excitation}

The longitudinal E-field (Fig. \ref{fig:butterfly}, left picture) shows maxima
of intensity along the streaming direction, as well as resonant excitations
consistent with the streaming electrons' pitch angle distribution, forming
cones of high intensity around the x-axis. An additional, but weaker component
perpendicular to the streaming direction is also visible.
The cones around the x-axis are the expected signature of beam-driven wave
excitation, following the beam direction and being widened by the electrons'
pitch-angle distribution. The components oblique and quasi-perpendicular to the
electron beams' direction are not explained by this mechanism, and are assumed
to be scattering products of three-wave interactions, as outlined in
\cite{KarlickyIAUS}.

\begin{figure}[htb] 
	\begin{center}
	\rotatebox{270}{\includegraphics[height=8cm]{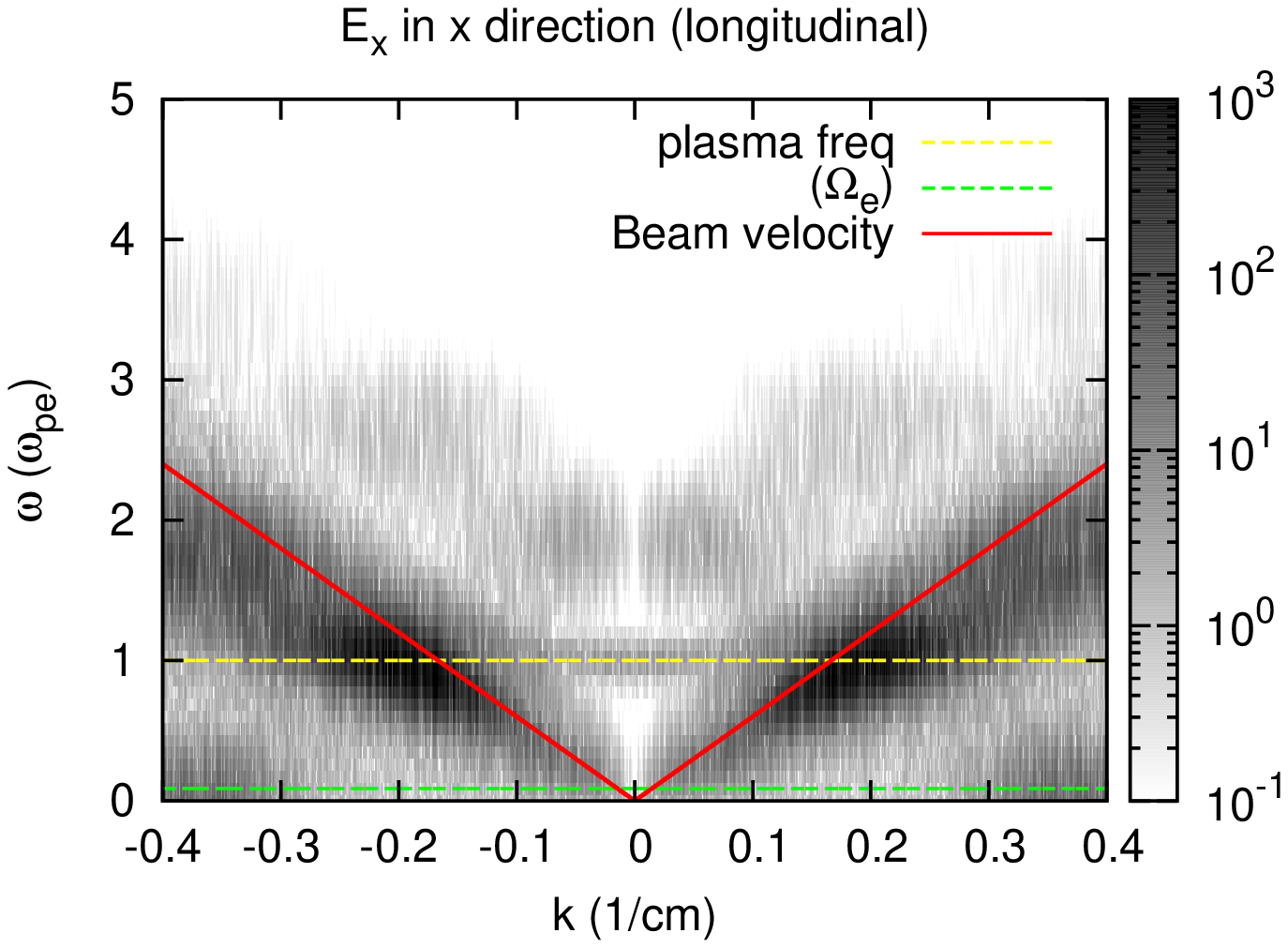}}
	\rotatebox{270}{\includegraphics[height=8cm]{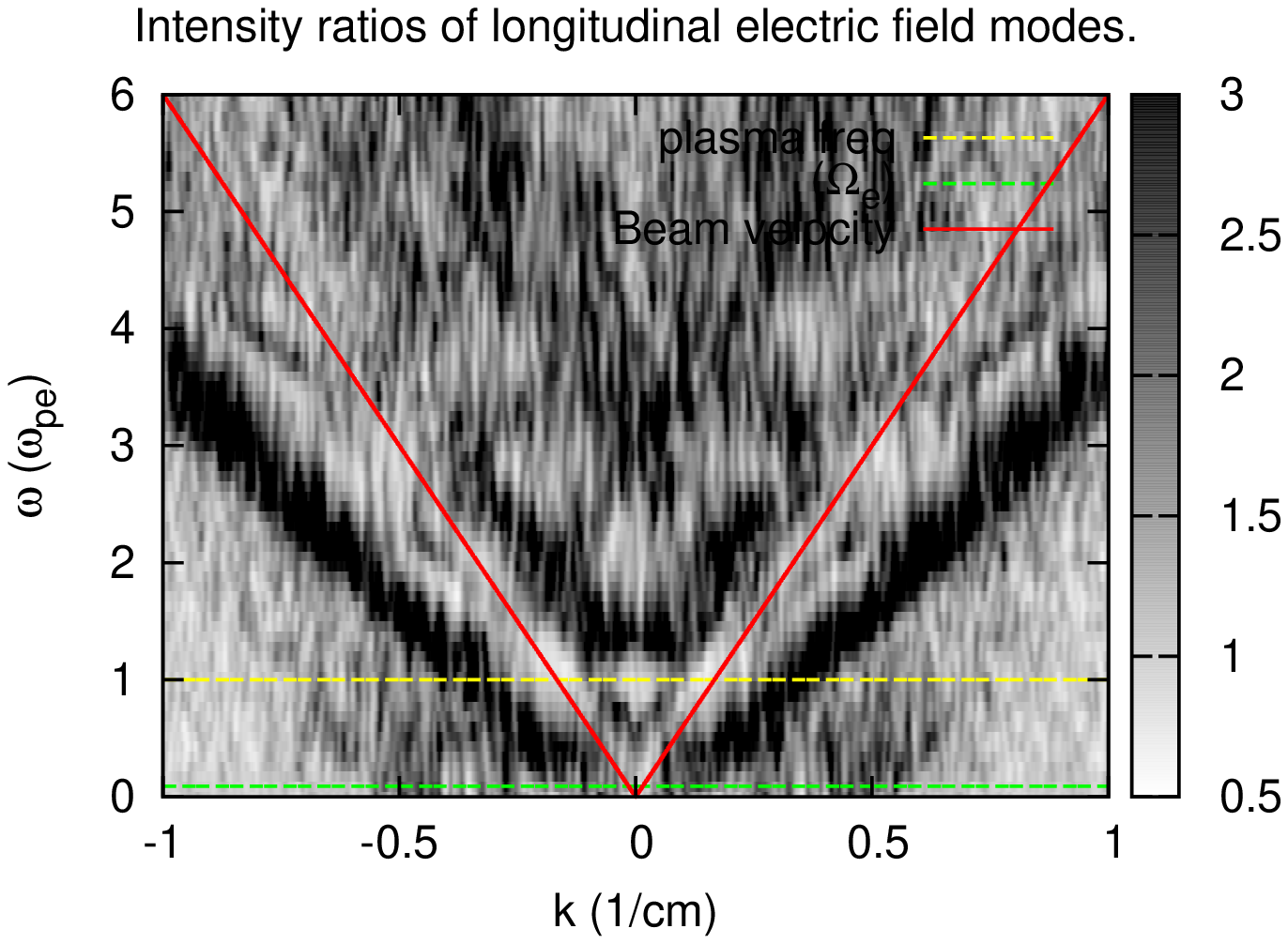}}
	\end{center}
	\caption{\textbf{Left:} Dispersion plot of the longitudinal electric field intensity along
	the beam direction. The beam resonance deposits large amounts of energy into
	the beam driven mode. \textbf{Right:} Increase (ratio) of electric field
	spectral energy between the weak beam and strong beam simulation.}
	\label{fig:dispersione}
\end{figure}

In Fig. \ref{fig:dispersione}, $k_y =0$ cuts through $\vec{k}$-space have been
taken for both the weak beam and strong beam case, and Fourier-transformed in
time to yield $k-\omega$ dispersion diagrams, allowing for identification of
individual wave modes. In the right picture, intensity ratios between the
weak beam and strong beam simulation are depicted, showing that the added wave
intensity primarily leads to a broadening of the beam mode excitation towards
lower frequencies.
The dominant contribution of energy in these plots is located on the beam mode,
with a constant phase velocity matching the average beam velocity
\citep{WillenGeneralizedLangmuir}. Around the plasma frequency, where the
beam-instability has its resonant maximum growth rate, the highest intensities are
observed. The two symmetric, counter-streaming beams lead to symmetric dispersion behaviour.

It should be noted that this beam mode is not an eigenmode of the background
plasma, but rather presents an entropy wave transported by the beam component.
As the beam weakens due to Landau damping, this mode has to undergo three-wave
interactions to decay into actual plasma eigenmodes, as theoretically predicted in
section \ref{theory}. Due to the short physical timescales of the kinetic
simulations conducted here, a relaxation of the electron beams' bump on tail has not
progressed to a large degree, so that the intensity in the beam mode is still
dominant at the end of the runs. This stands in contrast to in-situ satellite
observations, which are assumed to probe plasma conditions where most of the
beam mode's energy has already transferred into Langmuir waves. Conversion
efficiencies of beam to wave energies are therefore not directly comparable.

\subsection{Electromagnetic wave emission}
The $k$-space plot of the transverse magnetic field (Fig. \ref{fig:butterfly}, right)
shows two separate emission regions (and their symmetrical counterparts): one
at a low angle of about $20^\circ$ against the beam axis, the other at
near-perpendicular angles. Note that the signal-to-noise ratio of
these signals, being produced by higher-order processes, is much lower than in
the electric field components.

From theoretical considerations of momentum and energy conservation
(\ref{theory}), the near-perpendicular emission can be identified with the
harmonic emission process, whereas the lower-angle emission is consistent with
the fundamental frequency emission process.

\begin{figure}[htb]
	\begin{center}
		\rotatebox{270}{\includegraphics[height=8cm]{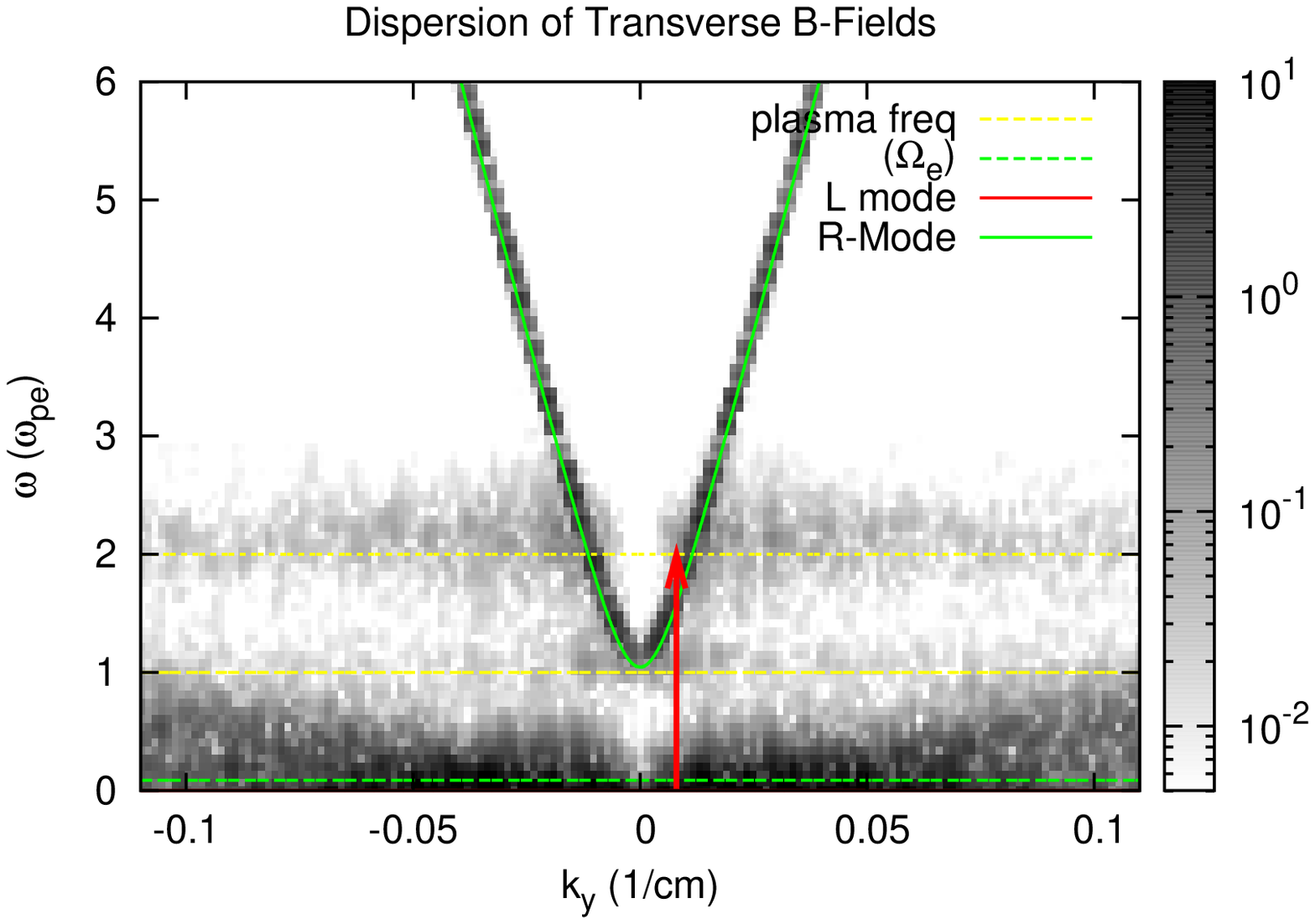}}
		\rotatebox{270}{\includegraphics[height=8cm]{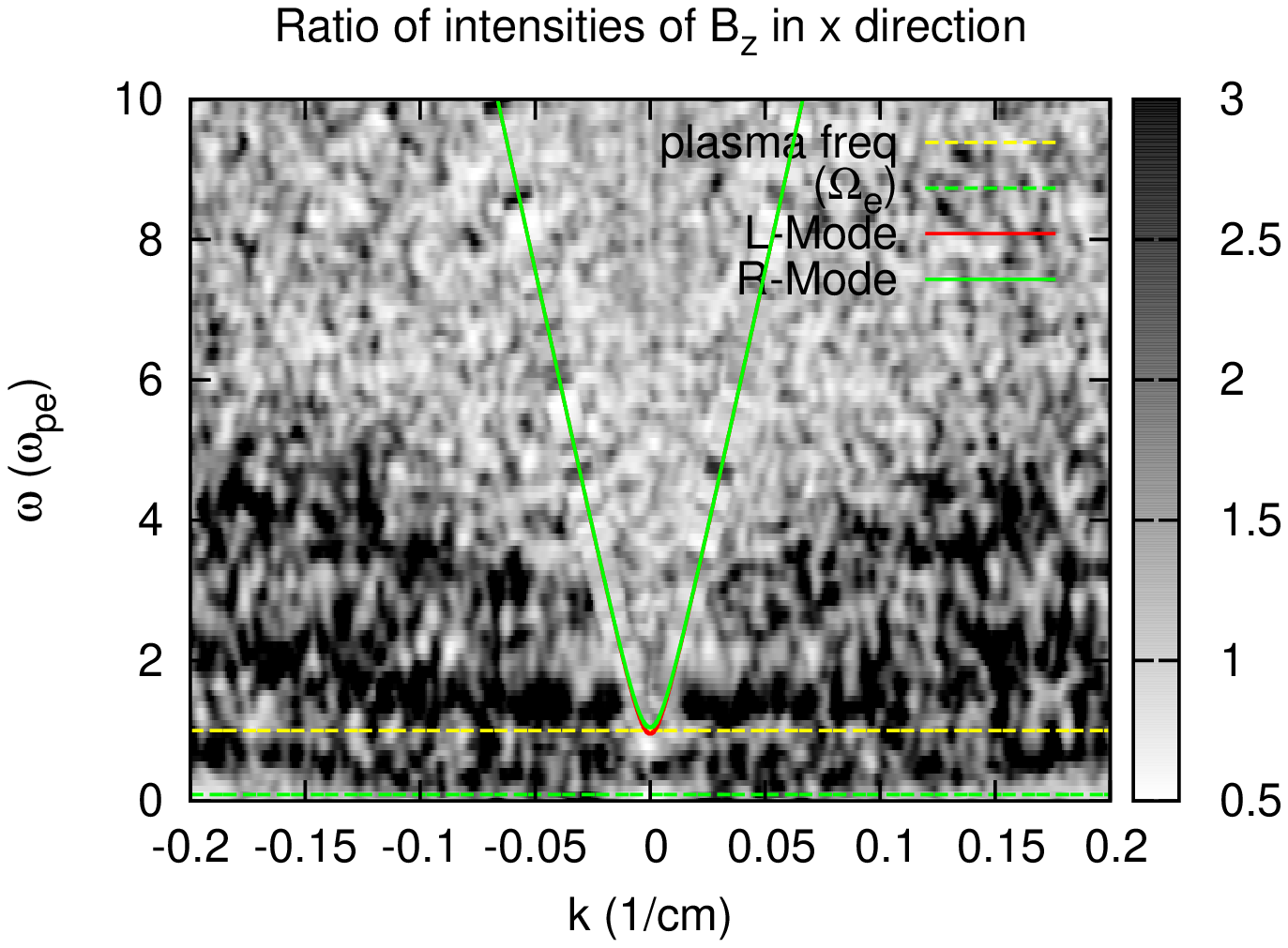}}
	\end{center}
	\caption{\textbf{Left:} Dispersion plot of the transverse magnetic field
	intensity (arb. units), with wave propagation direction perpendicular to the
	electron beam. Along the electromagnetic mode, resonances at $\omega_{pe}$ and $2
	\omega_{pe}$ are visible (dashed lines).
	The theoretically predicted point of interaction is denoted by the red arrow.
	\textbf{Right:} Ratio of intensities between the weak beam and strong beam
	simulations. Significant increase in energy is visible at the fundamental and
	harmonic frequency bands, especially at their upper edge.}
	\label{fig:dispersionb}
\end{figure}

The $k_x =0$ plane of $\vec{k}$-space was used to obtain the $k-\omega$
dispersion plot in Fig. \ref{fig:dispersionb} (left). In the plot, the electromagnetic
mode is clearly visible as a parabola with cutoff at the plasma frequency.
Additionally, lower velocity modes dominate the low $\omega$ values, probably
mainly consisting of improperly resolved Alfv\'en waves (which are of no
significance to the processes here).

Most importantly however, two horizontal bands in this plot are noticeable,
localized at $\omega = \omega_{pe}$ and $2 \omega_{pe}$ respectively. These
bands do not follow the dispersion relations of any mode predicted by linear
theory, which indicates that their origin is of non-linear nature.
Kinematically, they correlate with the predicted three-wave
interaction processes in Eq. \ref{lts} and \ref{llt} - assuming a coupling of
waves of opposite-directed $k$-vectors from the resonantly excited longitudinal
modes.

The $k$-value of maximum coupling of the harmonic emission predicted by Eq.\
\ref{analytic4} is $k_y = 8.3 \cdot 10^{-3} \mathrm{cm}$, which matches the
point of resonance visible in the simulation results (marked as a red arrow in
Fig. \ref{fig:dispersionb}).
It is noteworthy that the intensities of the two bands are quite similar, in
contrast to observations which typically show higher intensity in the harmonic
than fundamental emission. This is, however, assumed to be a propagation
effect: due to the close proximity of the emission to the plasma frequency, the
local medium has a much higher index of refraction for the fundamental than the
harmonic emission. The harmonic emission can hence escape the emission region
much easier, and leads to a stronger signal in observations.

Apart from the plasma frequency and its harmonic, some intensity is also
visible along the electromagnetic mode's dispersion relation at other
frequencies. This excitation stems from the inherent numerical noise of the
particle in cell simulation, caused by the relatively low particle number per
cell.

\section{Conclusion}

Using our fully relativistic particle-in-cell code ACRONYM, we have simulated
electron beam driven wave excitation and subsequent wave couplings in a
CME-foreshock environment. The complete spatial and temporal information
obtainable from a numerical simulation provides improved insight into the wave
processes in the emission region over satellite measurements:
Analysis of resulting k-space wave distributions
yields kinematically sound emission patterns, consistent with analytical
predictions about the type II emission mechanism. Further analysis of spatial
and temporal data provides evidence of nonlinear three-wave interaction between
beam driven electrostatic modes to create electromagnetic modes at fundamental
and harmonic frequencies with roughly equivalent intensities.

The fundamental and harmonic emission features visible in $k$-space
representations of the simulated fields match those produced in
\cite{KarlickyIAUS}. Since the emission in our runs was self-consistently
produced from electron beams exciting Langmuir waves instead of artificial
monochromatic excitation, these features are less strongly peaked, noisier
and less pronounced, but otherwise share the emission characteristics.

\subsection{Outlook}
The beam intensity is just one of multiple tunable parameters of the emission
process. Further parameters include beam velocity, background magnetic field
intensity and the beam electrons' pitch angle distribution. Simulations are
currently underway, studying the behaviour of the system under variation of these parameters.

The enormous computational demands of particle-in-cell simulations are still a
limiting factor, determining the maximum temporal extent of the runs. Slow wave
interaction processes, like the scattering of Langmuir waves on sound waves
(Eq. \ref{eq1}) are still insufficiently resolvable, and future improvements in
computing power and kinetic plasma simulation techniques will be required to
create complete, all-encompassing simulations of this emission mechanism.

An alternative simulation method worth studying would be a Vlasov code which,
while equally or even more numerically challenging, would be able to simulate
similar plasma behaviour free of effects due to statistical noise.

\section{Acknowledgements}
The simulations in this research have been made possible through computing grants by
the Juelich Supercomputing Centre (JSC) and the CSC - IT Center for Science Ltd., Espoo, Finland.\newline
UG acknowledges support by the Elite Network of Bavaria.\newline
FS acknowledges support from the Deutsche Forschungsgemeinschaft through grant SP 1124/1\newline
This work has been supported by the European Framework Programme 7 Grant
Agreement SEPServer - 262773\newline
We wish to express special thanks to our reviewer for his constructive and helpful comments.

\bibliographystyle{apj}
\bibliography{ursg}

\begin{thebibliography}{21}
\expandafter\ifx\csname natexlab\endcsname\relax\def\natexlab#1{#1}\fi

\bibitem[{{Aurass} {et~al.}(1994){Aurass}, {Klein}, \& {Mann}}]{Aurass94}
{Aurass}, H., {Klein}, K.-L., \& {Mann}, G. 1994, in ESA Special Publication,
  Vol. 373, Solar Dynamic Phenomena and Solar Wind Consequences, the Third SOHO
  Workshop, ed. {J.~J.~Hunt}, 95

\bibitem[{Cairns {et~al.}(2003)Cairns, Knock, Robinson, \& Kuncic}]{keyhere}
Cairns, I., Knock, S., Robinson, P., \& Kuncic, Z. 2003, Space Science Reviews,
  107, 27

\bibitem[{Cane {et~al.}(1987)Cane, Sheeley, \& Howard}]{Cane1987}
Cane, H.~V., Sheeley, J., \& Howard, R.~A. 1987, J. Geophys. Res., 92, 9869

\bibitem[{Ganse {et~al.}(2011{\natexlab{a}})Ganse, Kilian, Vainio, \&
  Spanier}]{UrsSolarPhysics}
Ganse, U., Kilian, P., Vainio, R., \& Spanier, F. 2011{\natexlab{a}}, Submitted
  to Solar Physics

\bibitem[{Ganse {et~al.}(2011{\natexlab{b}})Ganse, Spanier, \&
  Vainio}]{GanseIAUS}
Ganse, U., Spanier, F., \& Vainio, R. 2011{\natexlab{b}}, in IAU Symposium,
  Vol. 274, IAU Symposium, ed. {A.~Bonanno, E.~de Gouveia Dal Pino, \&
  A.~G.~Kosovichev}, 470--472

\bibitem[{{J. I. Sakai} \& {M. Karlick\'y}(2008)}]{SakaiBullshit}
{J. I. Sakai}, \& {M. Karlick\'y}. 2008, A\&A, 478, L15

\bibitem[{Karlicky \& Barta(2010)}]{KarlickyIAUS}
Karlicky, M., \& Barta, M. 2010, Proceedings of the International Astronomical
  Union, 6, 252

\bibitem[{{Karlick{\'y}} \& {Vandas}(2007)}]{KarlickyVandas}
{Karlick{\'y}}, M., \& {Vandas}, M. 2007, \planss, 55, 2336

\bibitem[{Kilian {et~al.}(2012)Kilian, Burkart, \& Spanier}]{acronym11}
Kilian, P., Burkart, T., \& Spanier, F. 2012, in High Performance Computing in
  Science and Engineering '11, ed. W.~E. Nagel, D.~B. Kröner, \& M.~M. Resch
  (Berlin Heidelberg: Springer), 5--13

\bibitem[{{Knock} {et~al.}(2001){Knock}, {Cairns}, {Robinson}, \&
  {Kuncic}}]{KnockModel}
{Knock}, S.~A., {Cairns}, I.~H., {Robinson}, P.~A., \& {Kuncic}, Z. 2001,
  Journal of Geophysical Research, 106, 25041

\bibitem[{{Melrose}(1986)}]{Melrose}
{Melrose}, D.~B. 1986, {Instabilities in Space and Laboratory Plasmas}
  (Instabilities in Space and Laboratory Plasmas, by D.~B.~Melrose,
  pp.~288.~ISBN 0521305411.~Cambridge, UK: Cambridge University Press)

\bibitem[{{Nelson} \& {Melrose}(1985)}]{NelsonMelrose}
{Nelson}, G.~J., \& {Melrose}, D.~B. 1985, {Type II bursts} (Cambridge
  University Press), 333--359

\bibitem[{Pulupa \& Bale(2008)}]{Pulupa2007}
Pulupa, M., \& Bale, S.~D. 2008, Astrophysical Journal, 676, 1330

\bibitem[{Reiner {et~al.}(1998)Reiner, Kaiser, Fainberg, \& Stone}]{reiner1998}
Reiner, M.~J., Kaiser, M.~L., Fainberg, J., \& Stone, R.~G. 1998, J. Geophys.
  Res., 103, 29651

\bibitem[{Schmidt \& Gopalswamy(2008)}]{schmidtCMEshocks}
Schmidt, J.~M., \& Gopalswamy, N. 2008, Journal of Geophysical Research, 113,
  A08104

\bibitem[{Spanier \& Vainio(2009)}]{SpanierVainio09}
Spanier, F., \& Vainio, R. 2009, Advanced Science Letters, 2, 337

\bibitem[{Thejappa {et~al.}(2012)Thejappa, MacDowall, \&
  Bergamo}]{Thejappa2012}
Thejappa, G., MacDowall, R.~J., \& Bergamo, M. 2012, \apj, 745, 187

\bibitem[{Wild \& McCready(1950)}]{WildMcCready}
Wild, J., \& McCready, L. 1950, Australian Journal of Scientific Research, 3,
  387

\bibitem[{{Willes} \& {Cairns}(2000)}]{WillenGeneralizedLangmuir}
{Willes}, A.~J., \& {Cairns}, I.~H. 2000, Physics of Plasmas, 7, 3167

\bibitem[{Yu \& Guangli(2008)}]{Guangli}
Yu, H., \& Guangli, H. 2008, Advances in Space Research, 41, 1202

\bibitem[{{Zlotnik} {et~al.}(1998){Zlotnik}, {Klassen}, {Klein}, {Aurass}, \&
  {Mann}}]{ThirdHarmonic}
{Zlotnik}, E.~Y., {Klassen}, A., {Klein}, K.-L., {Aurass}, H., \& {Mann}, G.
  1998, \aap, 331, 1087

\end{thebibliography}

\clearpage

\end{document}